\documentclass[12pt]{iopart}
\usepackage{iopams} 
\input epsf
\expandafter\let\csname equation*\endcsname\relax
 \expandafter\let\csname endequation*\endcsname\relax
 \usepackage[fleqn]{amsmath}
\usepackage{amsfonts}
\usepackage{amssymb}
\usepackage{amsopn}
\usepackage{epsfig}
\usepackage{graphics,psfrag,rotating}
\usepackage{graphicx}
\usepackage{dcolumn}
\usepackage{bm}
\usepackage{epstopdf}
\usepackage{color}
\usepackage[usenames,dvipsnames,svgnames]{xcolor}
\usepackage[colorlinks=true,
      linkcolor=red,
      urlcolor=gray,
      citecolor=blue]{hyperref}
 \usepackage{hyperref}

\def \D {\tilde{\nabla}}

\def \ep {\varepsilon}
\def\rd {\displaystyle{\cdot}}

\def\div {\mbox{div}\,}

\def \ts {\textstyle}
\def\om{\omega}

\def\3nab{\tilde{\nabla}}

\def\la {\langle}
\def\ra {\rangle}

\def\div{\mbox{div}}
\def\tl{\tilde}
\def\hsp5{\hspace{5mm}}
\newcommand{\sfrac}[2]{{\textstyle{#1\over#2}}}
\def\case#1/#2{\textstyle\frac{#1}{#2}}
\def\rd {\displaystyle{\cdot}}
\def\ts {\textstyle}
\def\ber {\begin{eqnarray}}
\def\eer {\end{eqnarray}}
\def\bea {\begin{eqnarray}}
\def\eea {\end{eqnarray}}

\def\ts {\textstyle}
\def\bc {\begin{center}}
\def\ec {\end{center}}
\def\case#1/#2{\frac{#1}{#2}}

\newcommand{\bw}{\begin{widetext}}
\newcommand{\ew}{\end{widetext}}

\newcommand{\be}{\begin{equation}}
\newcommand{\bse}{\begin{subequation}}
\newcommand{\ese}{\end{subequation}}
\newcommand{\ee}{\end{equation}}
\newcommand{\eei}{\end{eqnarray}\indent\indent}
\newcommand{\ba}{\begin{array}}
\newcommand{\ea}{\end{array}}
\newcommand{\bal}{\begin{eqnarray}}
\newcommand{\eal}{\end{eqnarray}}

\newcommand{\hs}{\,-\,}

\def\case#1/#2{\textstyle\frac{#1}{#2} }
\newcommand{\nb}{\nabla}

\newcommand{\bver}{\begin{verbatim}}
\newcommand{\ever}{\end{verbatim}}

\begin{document}
\title{Irrotational-fluid cosmologies in fourth-order gravity}
\author{Amare Abebe$^{1,2}$ and Maye Elmardi$^{3,4}$}
\address{$^{1}$Department of Physics, North-West University (NWU), Mahikeng 2735, South Africa}
\address{$^{2}$ Entoto Observatory and Research Center, P.O.Box 33679, Addis Ababa, Ethiopia}
\address{$^3$Astrophysics, Cosmology and Gravity Centre (ACGC), University of Cape Town, Rondebosch, 7701, South Africa}
\address{$^{4}$ Department of Mathematics and Applied Mathematics, University of Cape Town, Rondebosch, 7701, South Africa}
\date{\today}

%
%
\begin{abstract}

 In this paper, we explore classes of irrotational-fluid cosmological models in the context of $f(R)$-gravity in an attempt to put some theoretical and mathematical restrictions on the form of the $f(R)$ gravitational Lagrangian. In particular, we investigate the consistency of linearised dust models for shear-free cases as well as in the limiting cases when either the gravito-magnetic or gravito-electric components of the Weyl tensor vanish. We also discuss the existence and consistency of classes of non-expanding irrotational spacetimes in $f(R)$-gravity.\end{abstract}
 
\pacs{04.50.Kd, 04.25.Nx} \maketitle


\section{Introduction}
The recent discovery of the accelerated rate of cosmic expansion has inspired a wave of new research into the nature of gravitational physics. New alternatives and/or generalisations to Einstein's General Relativity (GR) theory abound already. One such generalisation comes in the form of {\it $f(R)$-gravity theories}. These are gravitational models based on the inclusion of higher-order curvature invariants in the Einstein-Hilbert action that result in fourth-order field equations \cite{sotiriou10, clifton12, capozziello11extended, modesto12, biswas12, nojiri03}.

Recent attempts to explain away early-universe cosmic inflation \cite{staro80} and late-time cosmic acceleration \cite{carroll04,faraoni08, sotiriou07,biswas12, nojiri2011unified, nojiri06, nojiri07, sriva08, capoz06, de2010f, magnano87} using $f(R)$-gravity have brought the latter under severe scrutiny. Although the nonlinearity of the field equations has complicated the analysis of the detailed physics of these theories, recent cosmological applications include studies on the dynamics of homogeneous cosmological models \cite{rippl96, amendola07, CDCT05,carloni2009, leach06, goheer2009power, dunsby10} and the linear growth of large-scale structures \cite{bean07,song07, abebe12, carloni08,ananda09, abebe13, ananda08, abebe14b, dombriz08}.

 In order to understand the dynamics of nonlinear fluid flows in $f(R)$ theories, it is important to understand the relationship between their Newtonian and general relativistic limits. This is relevant both in the physics of gravitational collapse and the late (nonlinear) stages of cosmic structure formation \cite{matarrese94, bert94, maartens98, wyll06, maartens94,ellis85id}. 
The differential properties of time-like geodesics describe the fluid flows in cosmology \cite{abebe12, ellis67, Ellis98}. The expansion $\Theta$, shear (distortion) $\sigma_{ab}$, rotation (vorticity) $\omega^a$, and acceleration $A_a$ of the four-velocity field $u^a$ tangent to the fluid flowlines describe kinematics of such fluid flows [as will be seen in Eqn \eref{congruence} shortly]. The generalised field equations (governing the fluid flows) are obtained by contracting the Ricci identities along and orthogonal to $u^a$.

Consistency analyses of the field equations for different models where integrability conditions, combinations of the Bianchi identities and their consequences, arise from imposing external restrictions have been made over the years \cite{maartens98,wyll06, maartens97, van97,lesame95, elst98, macc98}. The introduction of integrability conditions in $f(R)$ gravitational models, and their conservations should these conditions be constraints, provides useful insight into the forms of the $f(R)$ action. It also helps us explore the existence and nature of some universe models that would otherwise not exist under the stricter requirement of the action involving the Einstein-Hilbert Lagrangian of GR.

In this paper, we explore general properties of classes of irrotational-fluid spacetimes characterised by the vanishing of vorticity $\omega_a$, but generally non-vanishing energy density $\mu\;,$ shear and a locally free gravitational field which is covariantly described \cite{maartens98} by the gravito-electric (GE) and gravito-magnetic (GM) tensors, $E_{ab}$ and $H_{ab}\;,$ respectively. 

The rest of this paper is organised as follows: in Sec. \ref{covsec} we give a covariant description and the general linearised (with respect to the Friedmann-Lema\^itre-Robertson-Walker (FLRW) background) field equations of $f(R)$-gravity models. In Sec. \ref{irrotsec} we specialise to irrotational fluid spacetimes and analyse the properties of dust (in Sec. \ref{dustsubsec}) and non-expanding (in Sec. \ref{nonexpanding}) subclasses of spacetimes.
Finally in Sec. \ref{concsec} we discuss the results and give conclusions.


Natural units ($\hbar=c=k_{B}=8\pi G=1$)
will be used throughout this paper, and Latin indices run from 0 to 3.
The symbols $\nabla$, $\D$ and the overdot $^{.}$ represent the usual covariant derivative, the spatial covariant derivative, and differentiation with respect to cosmic time. We use the
$(-+++)$ spacetime signature and the Riemann tensor is defined by
\begin{eqnarray}
R^{a}_{bcd}=\Gamma^a_{bd,c}-\Gamma^a_{bc,d}+ \Gamma^e_{bd}\Gamma^a_{ce}-
\Gamma^f_{bc}\Gamma^a_{df}\;,
\end{eqnarray}
where the $\Gamma^a_{bd}$ are the Christoffel symbols (i.e., symmetric in
the lower indices) defined by
\begin{equation}
\Gamma^a_{bd}=\frac{1}{2}g^{ae}
\left(g_{be,d}+g_{ed,b}-g_{bd,e}\right)\;.
\end{equation}
The Ricci tensor is obtained by contracting the {\em first} and the
{\em third} indices of the Riemann tensor:
\begin{equation}\label{Ricci}
R_{ab}=g^{cd}R_{cadb}\;.
\end{equation}
Unless otherwise stated, primes $^{'}$ etc are shorthands for derivatives with respect to the Ricci scalar
\be
R=R^{a}{}_{a}\;
\ee
and $f$ is used as a shorthand for $f(R)$.
\section{Covariant Linearised Field Equations}\label{covsec}

The standard $f(R)$ gravitational action with a matter field contribution to the Lagrangian, ${\cal L}_m\;,$ is given by \cite{clifton12, capozziello11extended, de2010f, carloni08, buchdal70}
\be
{\cal A}= \sfrac12 \int d^4x\sqrt{-g}\left[f(R)+2{\cal L}_m\right]\;.
\label{action}
\ee
Using the variational principle of least action with respect to the metric $g_{ab}$,
the generalised Einstein Field Equations (EFEs) can be given in a compact form as
 \be
 G_{ab}=\tl T^{m}_{ab}+T^{R}_{ab}\equiv T_{ab}\;,
 \ee
 with 
 \be\label{emt}
\tl T^{m}_{ab}\equiv \frac{T^{m}_{ab}}{f'}\;,~~~~
T^{R}_{ab}\equiv \frac{1}{f'}\left[\sfrac{1}{2}(f-Rf')g_{ab}+\nb_{b}\nb_{a}f'-g_{ab}\nb_{c}
\nb^{c}f' \right]
\ee
defined as the effective matter and {\it curvature} (considered as a fluid) energy-momentum tensors (EMTs), respectively.
Standard matter has an EMT given by 
\be T^{m}_{ab} = \mu_{m}u_{a}u_{b} + p_{m}h_{ab}+ q^{m}_{a}u_{b}+ q^{m}_{b}u_{a}+\pi^{m}_{ab}\;,\ee
where $\mu_{m}$, $p_{m}$, $q^{m}_{a}$ and $\pi^m_{ab}$ are energy density, 
pressure, heat flux and anisotropic pressure respectively. Here $u^a\equiv \frac{dx^a}{dt}$ is the $4$-velocity of fundamental observers comoving with the fluid and is used to define the 
\textit{covariant time derivative} for any tensor 
${S}^{a..b}{}_{c..d} $ along an observer's worldlines:
\be
\dot{S}^{a..b}{}_{c..d}{} = u^{e} \nb_{e} {S}^{a..b}{}_{c..d}\;.
\ee
The projection tensor into the tangent 3-spaces orthogonal to $u^a$ is given by $ h_{ab}\equiv g_{ab}+u_au_b$
and is used to define the fully orthogonally 
\textit{projected covariant derivative} for any tensor ${S}^{a..b}{}_{c..d} $:
\be
\tl\nb_{e}S^{a..b}{}_{c..d}{} = h^a{}_f h^p{}_c...h^b{}_g h^q_d 
h^r{}_e \nb_{r} {S}^{f..g}{}_{p..q}\;,
\ee
with total projection on all the free indices. The orthogonally \textit{projected symmetric trace-free} (PSTF) part of vectors and rank-2 tensors is defined as
\be
V^{\langle a \rangle} = h^{a}_{b}V^{b}~, ~ S^{\langle ab \rangle} = \left[ h^{(a}{}_c {} h^{b)}{}_d 
- \sfrac{1}{3} h^{ab}h_{cd}\right] S^{cd}\;,
\label{PSTF}
\ee
and the volume element for the rest spaces orthogonal to $u^a$ is given by \cite{Ellis98, Abebe2011, betschart}
\be
\ep_{abc}=u^{d}\eta_{dabc}=-\sqrt{|g|}\delta^0{}_{\left[ a \right. }\delta^1{}_b\delta^2{}_c\delta^3{}_{\left. d \right] }u^d\Rightarrow \ep_{abc}=\ep_{[abc]},~\ep_{abc}u^{c}=0,
\ee
where $\eta_{abcd}$ is the 4-dimensional volume element with the properties
\be
\eta_{abcd}=\eta_{[abcd]}=2\ep_{ab[c}u_{d]}-2u_{[a}\ep_{b]cd}.
\ee
We define the covariant spatial divergence and curl of vectors and rank-2 tensors as \cite{maartens97}
\ber
&& \div V=\D^aV_a\,,~~~~~~(\div S)_a=\D^bS_{ab}\,, \\
&& curl V_a=\ep_{abc}\D^bV^c\,,~~ curl S_{ab}=\ep_{cd(a}\D^cS_{b)}{}^d \,.
\eer
The first covariant derivative of $u^a$ can be split into its
irreducible parts as
\be\label{congruence}
\nb_au_b=-A_au_b+\sfrac13h_{ab}\Theta+\sigma_{ab}+\ep_{a b c}\omega^c,
\ee
where $A_a\equiv \dot{u}_a$, $\Theta\equiv \tl\nb_au^a$, 
$\sigma_{ab}\equiv \tl\nb_{\langle a}u_{b \rangle}$ and $\omega^{a}\equiv\ep^{a b c}\tl\nb_bu_c$.
The {\it Weyl conformal curvature tensor} $C_{abcd}$ is defined as \cite{Ellis98, betschart}
\be
C^{ab}{}_{cd}=R^{ab}{}_{cd}-2g^{[a}{}_{[c}R^{b]}{}_{d]}+\frac{R}{3}g^{[a}{}_{[c}g^{b]}{}_{d]}
\ee
and can be split into its ``electric'' and ``magnetic'' parts, respectively, as
\be
E_{ab}\equiv C_{agbh}u^{g}u^{h},~~~~~~~H_{ab}=\sfrac{1}{2}\eta_{ae}{}^{gh}C_{ghbd}u^{e}u^{d}.
\ee
$E_{ab}$ and $H_{ab}$ represent the free gravitational field \cite{Ellis98}, enabling gravitational action at a distance (tidal forces and gravitational waves), and influence the motion of matter and radiation through the geodesic deviation for timelike and null vector fields, respectively. 

Cosmological quantities that vanish in the background spacetime are considered to be first-order and gauge-invariant by virtue of the Stewart-Walker 
lemma \cite{SW74}.

In a multi-component fluid universe filled with standard matter fields (dust, radiation, etc) and curvature contributions, the total energy density, isotropic and anisotropic pressures and heat flux are given, respectively, by \cite{ carloni08}
\be\label{totaltherm}
\mu\equiv\frac{\mu_{m}}{f'}+\mu_{R}\;,~~~\;p\equiv\frac{p_{m}}{f'}+p_{R}\;,~~~~~~\;\pi_{ab}\equiv\frac{\pi^{m}_{ab}}{f'}+\pi^{R}_{ab}\;,~~~
q_{a}\equiv \frac{q^{m}_{a}}{f'}+q^{R}_{a}\;,
\ee
where the linearised thermodynamic quantities for the  curvature fluid component are defined as
\ber
&&\label{mur}\mu_{R}=\frac{1}{f'}\left[\sfrac{1}{2}(Rf'-f)-\Theta f'' \dot{R}+ f''\tilde{\nabla}^{2}R \right]\;,\\
&&\label{pr}p_{R}=\frac{1}{f'}\left[\sfrac{1}{2}(f-Rf')+f''\ddot{R}+f'''\dot{R}^{2}+\sfrac{2}{3}\left( \Theta f''\dot{R}-f''\tilde{\nabla}^{2}R \right) \right]\;,\\
&&\label{qar}q^{R}_{a}=-\frac{1}{f'}\left[f'''\dot{R}\tilde{\nabla}_{a}R +f''\tilde{\nabla}_{a}\dot{R}-\sfrac{1}{3}f''\Theta \tilde{\nabla}_{a}R \right]\;,\\
&&\label{pir} \pi^{R}_{ab}=\frac{f''}{f'}\left[\tilde{\nabla}_{\langle a}\tilde{\nabla}_{b\rangle}R-\sigma_{ab}\dot{R}\right]\;.
\eer
 
Applying the $1+3$-covariant decomposition \cite{ellis89,ellis12}\hs in which a fundamental observer slices spacetime into time and space\hs on the Bianchi and Ricci identities
\be\label{biricci}
\nb_{[a}R_{bc]d}{}^{e}=0\;,~~~~
(\nb_{a}\nb_{b}-\nb_{b}\nb_{a})u_{c}=R_{abc}{}^{d}u_{d}\;
\ee
 for the total fluid 4-velocity $u^{a}$, the following linearised propagation (evolution) and constraint equations are obtained \cite{ carloni08, maartens98}:
\ber
&&\label{mue}\dot{\mu}_{m}=-(\mu_{m}+p_{m})\Theta-\tl\nb^{a}q^{m}_{a}\;,\\
&&\dot{\mu}_{R}=-(\mu_{R}+p_{R})\Theta+\frac{\mu_{m}f''}{f'^{2}}\dot{R}-\D^{a}q^{R}_{a}\;,\\
&&\label{ray}\dot{\Theta}=-\sfrac13 \Theta^2-\sfrac12(\mu+3p)+\tl\nb_aA^a\;,\\
&&\label{qe}\dot{q}^{m}_{a}=-\sfrac{4}{3}\Theta q^{m}_{a}-\mu_{m}A_{a}\;,\\
&&\label{qer}\dot{q}^{R}_{a}=-\sfrac{4}{3}\Theta q^{R}_{a}+\frac{\mu_{m}f''}{f'^{2}}\D_{a}R-\D_{a}p_{R}-\D^{b}\pi^{R}_{ab}\;,\\
&&\label{propom}\dot{\omega}_{a}=-\sfrac23\Theta\omega_{a}-\sfrac{1}{2}\ep_{abc}\tl\nb^{b}A^{c}\;,\\
&&\label{sig}\dot{\sigma}_{ab}=-\sfrac{2}{3}\Theta\sigma_{ab}-E_{ab}+\sfrac{1}{2}\pi_{ab}+\tl\nb_{\la a}A_{b\ra}\;,\\
&&\label{gep}\dot{E}_{ab}+\sfrac{1}{2}\dot{\pi}_{ab}=\ep_{cd\langle a}\tl\nb^{c}H_{b\rangle }^{d}-\Theta E_{ab}-\sfrac{1}{2}\left(\mu+p\right)\sigma_{ab}
-\sfrac{1}{2}\tl\nb_{\langle a}q_{b\rangle}-\sfrac{1}{6}\Theta\pi_{ab}\;,\\
&&\label{gmp}\dot{H}_{ab} =-\Theta H_{ab}-\ep_{cd\langle a}\tl\nb^{c}E_{b\rangle }^{d}+
\sfrac{1}{2}\ep_{cd\langle a}\tl\nb^{c}\pi^{~d}_{b\rangle}\;,
\eer

\ber
&&\label{R4} (C^{1})_{a}:=\D^{b}\sigma_{ab}-\sfrac{2}{3}\tl\nb_{a}\Theta+\ep_{abc}\tl\nb^{b}\omega^{c}+q_{a}=0\;,\\
&&\label{R6} (C^{2})_{ a b}:=\ep_{cd(a}\tl\nb^{c}\sigma_{b)}{}^{d}+\tl\nb_{\langle a}\omega_{b \rangle}-H_{a b}=0\;,\\
&&\label{B6} (C^{3})_{a}:=\tl\nb^{b}H_{ab}+(\mu+p)\omega_{a}+\sfrac12\ep_{abc}\tl\nb^{b}q^{c}=0\;,\\
&&\label{B5} (C^{4})_{a}:=\tl\nb^{b}E_{ab}+\sfrac{1}{2}\tl\nb^{b}\pi_{ab}-\sfrac13\tl\nb_{a}\mu+
\sfrac13\Theta q_{a}=0\;,\\
&& \label{R5} (C^{5}):=\tl\nb^a\omega_a=0\;,\\
&&\label{B3} (C^{6})_{a}:=\tl\nb_{a}p_{m} +(\mu_{m}+p_{m}) A_{a}=0\;.
\eer
The evolution equations
propagate consistent initial data on some initial ($t=t_{0}$) hypersurface $S_{0}$ uniquely along the (generally future-directed)
reference timelike congruence whereas the constraints restrict the initial data to be specified. For consistency, the constraint equations must remain satisfied on any hypersurface $S_{t}$ for all comoving time $t\;.$ 

\section{Irrotational Spacetimes}\label{irrotsec}
Irrotational fluid flows admit geodesic timelike congruences with vanishing vorticity
\be
\label{Can}\om_{a}=0\;
\ee
and characterise potential cosmological models for the late universe and gravitational collapse.
For a barotropic irrotational matter fluid with the equation of state $p_m=w\mu_m$, the evolution equations \eref{mue}-\eref{gmp} for this class of spacetimes can be rewritten as:
\ber
&&\label{mue2}\dot{\mu}_{m}=-(1+w)\mu_{m}\Theta-\tl\nb^{a}q^{m}_{a}\;,\\
&&\label{murdot}\dot{\mu}_{R}=-(\mu_{R}+p_{R})\Theta+\frac{\mu_{m}f''}{f'^{2}}\dot{R}-\D^{a}q^{R}_{a}\;,\\
&&\label{ray2}\dot{\Theta}=-\sfrac13 \Theta^2-\sfrac{1}{2f'}(1+3w)\mu_{m}-\sfrac12\left(\mu_{R}+3p_{R}\right)+\tl\nb_aA^a\;,\\
&&\label{qe2}\dot{q}^{m}_{a}=-\sfrac{4}{3}\Theta q^{m}_{a}-\mu_{m}A_{a}\;,\\
&&\label{qardot}\dot{q}^{R}_{a}=-\sfrac{4}{3}\Theta q^{R}_{a}+\frac{\mu_{m}f''}{f'^{2}}\D_{a}R-\D_{a}p_{R}-\D^{b}\pi^{R}_{ab}\;,\\
&&\label{sig2}\dot{\sigma}_{ab}=-\sfrac{2}{3}\Theta\sigma_{ab}-E_{ab}+\sfrac{1}{2}\pi_{ab}+\tl\nb_{\la a}A_{b\ra}\;,\\
&&\label{gep2}\dot{E}_{ab}+\sfrac{1}{2}\dot{\pi}_{ab}=\ep_{cd\langle a}\tl\nb^{c}H_{b\rangle }^{d}-\Theta E_{ab}-\sfrac{1}{2}\left(\mu+p\right)\sigma_{ab}
-\sfrac{1}{2}\tl\nb_{\langle a}q_{b\rangle}-\sfrac{1}{6}\Theta\pi_{ab}\;,\\
&&\label{gmp2}\dot{H}_{ab} =-\Theta H_{ab}-\ep_{cd\langle a}\tl\nb^{c}E_{b\rangle }^{d}+
\sfrac{1}{2}\ep_{cd\langle a}\tl\nb^{c}\pi^{~d}_{b\rangle}\;,
\eer
and are constrained by the following equations:
\ber
&&\label{R4ir} (C^{1\ast})_{a}:=\D^{b}\sigma_{ab}-\sfrac{2}{3}\tl\nb_{a}\Theta+q_{a}=0\;,\\
&&\label{R6ir} (C^{2\ast})_{ a b}:=\ep_{cd(a}\tl\nb^{c}\sigma_{b)}{}^{d}-H_{a b}=0\;,\\
&&\label{B6ir} (C^{3\ast})_{a}:=\tl\nb^{b}H_{ab}+\sfrac12\ep_{abc}\tl\nb^{b}q^{c}=0\;,\\
&&\label{B5ir} (C^{4\ast})_{a}:=\tl\nb^{b}E_{ab}+\sfrac{1}{2}\tl\nb^{b}\pi_{ab}-\sfrac13\tl\nb_{a}\mu+
\sfrac13\Theta q_{a}=0\;,\\
&&\label{B3ir} (C^{5\ast})_{a}:=w\tl\nb_{a}\mu_{m} +(1+w) \mu_{m}A_{a}=0\;,\\
&& \label{R5ir} (C^{6\ast})_{a}:=\ep_{abc}\tl\nb^{b}A^{c}=0\implies A_{a}=\D_{a}\psi \mbox{~for~some~scalar~} \psi\;.
\eer
The new constraint \eref{R5ir} arises as a result of our irrotational restriction. To check for temporal consistency, we propagate this constraint to obtain
\be
\left(\ep_{abc}\tl\nb^{b}A^{c}\right)^{.}=0\;,
\ee
which is an identity.
On the other hand, taking the curl of this constraint, one obtains
\ber
curl(curl(A_{a}))&&=\D_{a}\left(\D^{b}A_{b}\right)-\D^{2}A_{a}+\sfrac{2}{3}\left(\mu-\sfrac{1}{3}\Theta^{2}\right)A_{a}\\
&&=\D_{a}\left(\D^{b}\D_{b}\psi\right)-\D^{2}\D_{a}\psi+\sfrac{2}{3}\left(\mu-\sfrac{1}{3}\Theta^{2}\right)\D_{a}\psi\\
&&=\D_{a}\left(\D^{2}\psi\right)-\D^{2}\left(\D_{a}\psi\right)+\sfrac{2}{3}\left(\mu-\sfrac{1}{3}\Theta^{2}\right)\D_{a}\psi=0\;,
\eer
which is another identity by virtue of Eqns \eref{a19} and \eref{curlcurla}.
\subsection{Dust Spacetimes}\label{dustsubsec}
Pure dust spacetimes are characterised by \be w=0= p_{m}\;, q^{m}_{a}=0=A_{a}\;, \pi^{m}_{ab}=0\;,\ee and the linearised evolution and constraint equations read:
\ber
&&\label{mue2d}\dot{\mu}_{d}=-\mu_{d}\Theta\;,\\
&&\label{murdotd}\dot{\mu}_{R}=-(\mu_{R}+p_{R})\Theta+\frac{\mu_{d}f''}{f'^{2}}\dot{R}-\D^{a}q^{R}_{a}\;,\\
&&\label{ray2d}\dot{\Theta}=-\sfrac13 \Theta^2-\sfrac{1}{2f'}\mu_{d}-\sfrac{1}{2}\left(\mu_{R}+3p_{R}\right)\;,\\
&&\label{qardotd}\dot{q}^{R}_{a}=-\sfrac{4}{3}\Theta q^{R}_{a}+\frac{\mu_{d}f''}{f'^{2}}\D_{a}R-\D_{a}p_{R}-\D^{b}\pi^{R}_{ab}\;,\\
&&\label{sig2d}\dot{\sigma}_{ab}=-\sfrac{2}{3}\Theta\sigma_{ab}-E_{ab}+\sfrac{1}{2}\pi^{R}_{ab}\;,\\
&&\label{gep2d}\dot{E}_{ab}+\sfrac{1}{2}\dot{\pi}^{R}_{ab}=\ep_{cd\langle a}\tl\nb^{c}H_{b\rangle }^{d}-\Theta E_{ab}-\sfrac{1}{2}\left(\frac{\mu_{d}}{f'}+\mu_{R}+p_{R}\right)\sigma_{ab}
-\sfrac{1}{2}\tl\nb_{\langle a}q^{R}_{b\rangle}-\sfrac{1}{6}\Theta\pi^{R}_{ab}\;,\\
&&\label{gmp2d}\dot{H}_{ab} =-\Theta H_{ab}-\ep_{cd\langle a}\tl\nb^{c}E_{b\rangle }^{d}+
\sfrac{1}{2}\ep_{cd\langle a}\tl\nb^{c}\pi^{R~d}_{b\rangle}\;,
\eer

\ber
&&\label{R4irds} (C^{1d})_{a}:=\D^{b}\sigma_{ab}-\sfrac{2}{3}\tl\nb_{a}\Theta+q^{R}_{a}=0\;,\\
&&\label{R6irds} (C^{2d})_{ a b}:=\ep_{cd(a}\tl\nb^{c}\sigma_{b)}{}^{d}-H_{a b}=0\;,\\
&&\label{B6irds} (C^{3d})_{a}:=\tl\nb^{b}H_{ab}+\sfrac12\ep_{abc}\tl\nb^{b}q^{c}_{R}=0\;,\\
&&\label{B5irds} (C^{4d})_{a}:=\tl\nb^{b}E_{ab}+\sfrac{1}{2}\tl\nb^{b}\pi^{R}_{ab}-\sfrac{1}{3f'}\tl\nb_{a}\mu_{m}-\sfrac13\tl\nb_{a}\mu_{R}+
\sfrac13\Theta q^{R}_{a}=0\;.
\eer
Notice here that no new constraints appear.
\subsubsection{Shear-free spacetimes}
Over the years, the role of shear in GR and the special nature of shear-free cases in particular have been studied \cite{ellis67, godel52,goldberg62,robinson63,abebe2013}.
G\"{o}del showed \cite{godel52} that shear-free time-like geodesics of some spatially homogeneous universes cannot expand and rotate simultaneously and this result was later generalized by Ellis \cite{ellis67} to include inhomogeneous cases of shear-free time-like geodesics. Goldberg and Sachs, on the other hand, showed \cite{goldberg62} that shear-free null geodesic congruences {\it in vacuo} require an algebraically special Weyl tensor, a result later generalized by Robinson and Schild \cite{robinson63} to include non-vanishing, but special, forms of the Ricci tensor.

An interesting aspect of these shear-free solutions is that they do not hold in Newtonian gravitation theory \cite{ellis2011S, narlikar99, narlikar63} although Newtonian theory is a limiting case of GR under special circumstances, namely at low-speed relative motion of matter with no gravito-magnetic effects (vanishing magnetic part of the Weyl tensor) and hence no gravitational waves. 

Now if we turn off the shear, i.e., if we set
\be
\sigma_{ab}=0
\ee
in the above propagation equations, we get Eqn \eref{sig2d} turning into a new constraint
\be
\label{shearconst1} (C^{5d})_{a b}:=E_{ab}-\sfrac{1}{2}\pi^{R}_{ab}=0\;,\ee
the temporal and spatial consistencies of which have to be checked. It is interesting to note here that, unlike for shear-free dust spacetimes in GR, the electric component of the Weyl tensor does not vanish because of the non-vanishing contribution of the anisotropic pressure $\pi^R_{ab}$. 
However, we see from Eqn \eref{R6irds} that $H_{ab}$ identically vanishes, thus resulting in another constraint from Eqn \eref{gmp2d}:
\be
\label{shearconst2} (C^{6d})_{a b}:=\ep_{cd\langle a}\tl\nb^{c}E_{b\rangle }^{d}-
\sfrac{1}{2}\ep_{cd\langle a}\tl\nb^{c}\pi^{d~d}_{b\rangle}=0\;,
\ee
which is an identity by virtue of Eqn \eref{shearconst1}. 

From \eref{B6irds}, we see that $q^{R}_{a}$ is irrotational, and therefore can be written as the gradient of a scalar: \be\label{qairr}
q^{R}_{a}=\D_{a}\phi\;.\ee 
However, we already know from \eref{R4irds} that $q^{R}_{a}=\sfrac{2}{3}\tl\nb_{a}\Theta$. One can therefore conclude that in irrotational and shear-free dust spacetimes,
\be\label{qaph}
\phi=\sfrac{2}{3}\Theta +C\;,
\ee
for some spatially constant scalar $C$. We can rewrite this dynamical constraint on the expansion history, using Eqn \eqref{qar} in \eqref{qairr}, as
\be
\sfrac{2}{3}f'\D_{a}\Theta+\left(f''\dot{R}-\sfrac{1}{3}\Theta f''\right)\D_{a}R+f''\D_{a}\dot{R}=0\;.
\ee
For GR, i.e., $f=R\;,f'=1\;, f''=f'''=0\;,$ we obtain a spatially constant expansion since
\be
\D_a \Theta=0\;,
\ee
which is trivially true for the class of models under consideration.
 
Now to check for temporal consistency of Eqn \eref{shearconst1}, we take the time derivative of both sides of this equation to obtain the relation
\be
\dot{\pi}^{R}_{ab}+\sfrac{2}{3}\Theta \pi^{R}_{ab}-\sfrac{1}{2}\D_{\la a}q^{R}_{b\ra}=0\;,
\ee
which, using $q^{R}_{a}$ and $\pi^{R}_{ab}$ as defined by Eqns \eref{qar} and \eref{pir}, can be rewritten as
\be\label{tempcon}
\left[\frac{3}{2}\left(\frac{f'''}{f'}-\frac{f''^{2}}{f'^{2}}\right)\dot{R}-\frac{\Theta f''}{6f'}\right]\D_{\la a}\D_{b\ra}R+\sfrac{3f''}{2f'}\D_{\la a}\D_{b\ra}\dot{R}=0\;.
\ee
Thus irrotational shear-free dust spacetimes governed by $f(R)$ gravitational physics evolve consistently if Eqn \eref{tempcon} is satisfied. Note that Eqn \eref{tempcon} becomes an identity in the GR limit.

Now taking the curl of the above equation, we obtain
\be\label{spcon}
\left[\frac{3}{2}\left(\frac{f'''}{f'}-\frac{f''^{2}}{f'^{2}}\right)\dot{R}-\frac{\Theta f''}{6f'}\right]\ep_{cda}\D^{c}\D_{\la b}\D^{d\ra}R+\sfrac{3f''}{2f'}\ep_{cda}\D^{c}\D_{\la b}\D^{d\ra}\dot{R}=0\;,
\ee
which is an identity by virtue of Eqn \eref{a2}. Thus the temporal consistency condition given by Eqn \eref{tempcon} is satisfied on any initial hypersurface. Moreover, from Eqns \eref{shearconst2} and \eref{spcon}, we can conclude that {\it all irrotational shear-free dust spacetimes in $f(R)$-gravity are spatially consistent}.

A further restriction one can make for such (shear-free) spacetimes is turning off $E_{ab}$. This is a case of vanishing Weyl tensor (since $H_{ab}=0$ by virtue of Eqn \eqref{R6irds}), resulting in a locally conformally flat metric. For such a class of cosmological models Eqn \eref{gep2d} changes to the (linearised) constraint
\be\label{qdiv}
\tl\nb_{\langle a}q^{R}_{b\rangle}=0=\frac{1}{f'}\left[\left(\dot{R}f'''-\sfrac{1}{3}\Theta f''\right)\D_{\la a}\D_{b\ra}R+f''\D_{\la a}\D_{b\ra}\dot{R}\right]\;.
\ee 

Eqn \eref{sig2d} implies
\be 
\pi^{R}_{ab}=0=\frac{f''}{f'}\D_{\la a}\D_{b\ra}R\;,
\ee
which, for $f''\neq 0$, leads to the conclusion that
\be
\D_{\la a}\D_{b\ra}R=0\;.
\ee
Using this and the relation \eref{a0}, we see that Eqn \eref{qdiv} becomes an identity.
Thus the \emph{linearised $f(R)$ field equations in irrotational and shear-free dust spacetimes with vanishing Weyl tensor are consistent}.
\subsubsection{Dust solutions with $div H=0$}
The vanishing of the divergence of a non-zero $H_{ab}$ is a necessary condition for gravitational radiation \cite{lesame96, maar97c, maartens97lin}. Here we analyse the consistency of divergence-free GM ($\D^{b}H_{ab}=0$) scenarios in an effort to understand the nature of gravitationally radiating irrotational dust spacetimes.

We see that there are no new constraints arising as a result of imposing a divergence-free $H_{ab}$ to the field equations, but as in the shear-free case discussed above, Eqn \eref{B6irds} implies that $q^{R}_{a}$ satisfies Eqn \eref{qairr} whereas Eqn \eref{qaph} generalizes to
\be
\D_a\phi =\sfrac{2}{3}\D_a\Theta-\D^b\sigma_{ab}\;.
\ee 
 
Another interesting subclass of these models arises if both $H_{ab}$ and $E_{ab}$ are divergence-free. Described as ``purely radiative'' dust spacetimes \cite{van07}, such models should satisfy the additional modified constraint
\be
\tl\nb_{a}\mu_{m}+f'\tl\nb_{a}\mu_{R}+f'\Theta q^{R}_{a}-\sfrac{3f'}{2}\tl\nb^{b}\pi^{R}_{ab}=0
\ee 
as a result of Eqn \eqref{B5irds}. One can see from this result that purely radiative irrotational dust spacetimes in GR (f=R) should be spatially homogeneous (with $\tl\nb_{a}\mu_{m}=0$).

The so-called {\it Newtonian-like} spacetimes are described by the vanishing of the GM component of the Weyl tensor \cite{maartens1998}. Thus if the Weyl tensor is to have a purely GE component, then we notice from Eqn \eref{R6irds} that a curl-free shear is required, i.e., 
\be
H_{ab}= 0\implies \ep_{cd(a}\tl\nb^{c}\sigma_{b)}{}^{d} =0 \;,
\ee
 and the constraint \eref{shearconst2} is obtained from Eqn \eref{gmp2d}. Although such models are known to be of limited  applicability in GR-based cosmology, there are some interesting features in $f(R)$ cosmologies (Abebe et al, {\it in preparation}).
\subsubsection{Purely gravito-magnetic spacetimes}
These are models with vanishing gravito-electric component of the Weyl tensor and are referred to as {\it anti-Newtonian}\footnote{An earlier use of the word `anti-Newtonian' exists \cite{tomita72, tomita75}, where the word is used to refer to an earlier stage of the Universe when the dimension of irregularity exceeds the cosmological (Hubble) horizon.} models because they are considered to be the most extreme of non-Newtonian gravitational models \cite{maartens1998}. The only anti-Newtonian solutions in GR are the FLRW spacetimes \cite{wyll06,maartens1998}, but a recent covariant consistency analysis \cite{amare14a} has shown that linearised anti-Newtonian universes are permitted by some models of $f(R)$ gravity.

As can be seen from the set of equations \eref{mue2d}-\eref{B5irds}, no new constraint equations arise as a result of vanishing $E_{ab}$. This is because of the non-vanishing of $\pi^{R}_{ab}$ for generic $f(R)$ models; but in the GR limiting case Eqn \eref{gep2d} would have turned into a new constraint since $\pi^{R}_{ab}=0$.
\subsection{Non-expanding Spacetimes}\label{nonexpanding}
In this section we explore theoretical cases where the background spacetime is not expanding, i.e., $\Theta=0$ to analyse the kind of [in]consistencies one would obtain if  a universe with such properties existed. In this very special case, the linearised evolution equations \eref{mue2}-\eref{gmp2} become
\ber
&&\label{mue2s}\dot{\mu}_{m}=-\tl\nb^{a}q^{m}_{a}\;,\\
&&\label{qe2s}\dot{q}^{m}_{a}=\frac{w}{1+w}\D_{a}\mu_{m}\;,\\
&&\label{murdots}\dot{\mu}_{R}=\frac{\mu_{m}f''}{f'^{2}}\dot{R}-\D^{a}q^{R}_{a}\;,\\
&&\label{qardots}\dot{q}^{R}_{a}=\frac{\mu_{m}f''}{f'^{2}}\D_{a}R-\D_{a}p_{R}-\D^{b}\pi^{R}_{ab}\;,\\
&&\label{sig2s}\dot{\sigma}_{ab}=-E_{ab}+\sfrac{1}{2}\pi_{ab}+\tl\nb_{\la a}A_{b\ra}\;,\\
&&\label{gep2s}\dot{E}_{ab}+\sfrac{1}{2}\dot{\pi}_{ab}=\ep_{cd\langle a}\tl\nb^{c}H_{b\rangle }^{d}-\sfrac{1}{2}\left(\mu+p\right)\sigma_{ab}
-\sfrac{1}{2}\tl\nb_{\langle a}q_{b\rangle}\;,\\
&&\label{gmp2s}\dot{H}_{ab} =-\ep_{cd\langle a}\tl\nb^{c}E_{b\rangle }^{d}+
\sfrac{1}{2}\ep_{cd\langle a}\tl\nb^{c}\pi^{~d}_{b\rangle}\;,
\eer
whereas the revised constraint equations are given by

\ber
&&\label{R4irs} (C^{1s})_{a}:=\D^{b}\sigma_{ab}+q_{a}=0\;,\\
&&\label{R6irs} (C^{2s})_{ a b}:=\ep_{cd(a}\tl\nb^{c}\sigma_{b)}{}^{d}-H_{a b}=0\;,\\
&&\label{B6irs} (C^{3s})_{a}:=\tl\nb^{b}H_{ab}+\sfrac12\ep_{abc}\tl\nb^{b}q^{c}=0\;,\\
&&\label{B5irs} (C^{4s})_{a}:=\tl\nb^{b}E_{ab}+\sfrac{1}{2}\tl\nb^{b}\pi_{ab}-\sfrac13\tl\nb_{a}\mu=0\;,\\
&&\label{B3irs} (C^{5s})_{a}:=w\tl\nb_{a}\mu_{m} +(1+w) \mu_{m}A_{a}=0\;,\\
&&\label{ray2s} (C^{6s}):=\tl\nb_aA^a-\sfrac{1}{2f'}(1+3w)\mu_{m}-\sfrac12\left(\mu_{R}+3p_{R}\right)=0\;.
\eer
Eqn \eref{qe2s} has been obtained by using Eqn \eref{B3ir} into \eref{qe2} whereas Eqn \eref{ray2s} arises from Eqn \eref{ray2}, showing that in the non-expanding case the Raychaudhuri (acceleration) equation changes into a constraint.
\subsubsection{Dust Solutions}

In the case of dust ($A_{a}=0=q^{m}_{a}$), the active gravitational mass $\mu+3p=0$ because of Eqn \eref{ray2s}. Since \eref{mue2s} implies $\mu_{d}(t)=\mbox{const[ant]}$, we notice that \be\label{const2} \mu_{R}+3p_{R}=\mbox{const}\ee as well.
From the definitions \eref{mur} and \eref{pr} for $\mu_{R}$ and $p_{R}$ and the {\it trace equation}
\be\label{trace}
3f''\ddot{R}+3\dot{R}^2f'''+3\Theta\dot{R}f''-3f''\D^2R-Rf'+2f-\mu_m+3p_{m}=0\;,
\ee
we conclude that \eref{const2} implies\footnote{In GR, where $f(R)=R$, this translates into stating the obvious result that a constant $\mu_{d}$ implies a constant $R$ since $f''=0$.}
\be\label{fconstr}
f-2f''\D^{2}R=\mbox{const}\;.
\ee 
Thus any nonrotating and noexpanding dust spacetime in $f(R)$ cosmology should have a gravitational Lagrangian that satisfies Eqn \eqref{fconstr}.
\subsubsection{Shear-free Solutions}
As mentioned in the previous subsection, shear-free assumptions in cosmology result in many interesting (and at times intriguing) properties. Despite the limited applicability of such assumptions in standard cosmology\hs not least because the Universe is known, beyond any reasonable doubt, to be expanding\hs it is interesting to explore the different mathematical constraints one obtains if $f(R)$  is the basic gravitational physics behind such cosmology.

  If we make the shear-free assumption, the propagation equation \eref{sig2s} turns into the constraint

\be
\label{sig2sc} (C^{7s})_{ab}:=E_{ab}-\sfrac{1}{2}\pi_{ab}-\tl\nb_{\la a}A_{b\ra}\;,
\ee
whereas the constraint equations \eref{R4irs} and \eref{R6irs} imply $q_{a}=0$ and $H_{ab}=0$. This means that Eqn \eref{gep2s} reduces to
\be
\label{gep2ss}\dot{E}_{ab}+\sfrac{1}{2}\dot{\pi}_{ab}=0\;.
\ee
If we differentiate the new constraint \eref{sig2sc} with respect to cosmic time, and solve simultaneously with Eqn \eref{gep2ss} we obtain
\be
\label{gep2s2}\dot{E}_{ab}-(\tl\nb_{\la a}A_{b\ra})^{.}=0\;.
\ee
On the other hand, if we take the gradient of \eref{sig2sc} and solve simultaneously with \eref{B5irs}, we obtain
\be
\tl\nb^{b}E_{ab}-\sfrac16\tl\nb_{a}\mu-\sfrac12\D^b\D_{\la a}A_{b\ra}=0\;.
\ee
Moreover, the curl condition of \eref{sig2sc} is identically satisfied by virtue of Eqns \eref{gmp2s}, \eref{R5ir} and \eref{a2}. 

If we consider the special case of dust ($A_a=0=q^m_a\;, \pi^m_{ab}=0$) in this shear-free setting, then \eref{gep2s2} implies $E_{ab}=\mbox{const}$ in time. However, since $E_{ab}$ is related to $\pi^R_{ab}$ via Eqn \eref{sig2sc}, then $\pi^R_{ab}=\mbox{const}$ as well. This dictates that, because of \eref{pir}, the term 
$\frac{f''}{f'}\D_{\la a}\D_{b\ra}R$ be constant in cosmic time. It is also no coincidence that \eref{fconstr} is recovered for this subclass as a result of \eref{ray2s}.

Another interesting point to note about non-expanding, shear-free dust spacetimes is that since $q_a=0\implies q^R_a=0$, we are dictated by Eqn \eref{qar} to conclude:
\be\label{nesfd}
\left(f''\tilde{\nabla}_{a}R\right)^{.}=0\;,
\ee 
thus putting a constraint on the form of the viable $f(R)$ gravitational action that describes such spacetimes. We notice that Eqn \eqref{nesfd} is an identity in GR.
\section{Discussions and Conclusion}\label{concsec}
A completely general covariant analysis of irrotational fluids in $f(R)$ cosmology requires taking nonlinear effects into account. This paper is meant to be a first step in that direction, and we have looked at the limits of irrotational fluid spacetimes for some specialised solutions linearised around the FLRW background. 

We have shown that the only constraint arising as a result of the irrotational fluid assumption in $f(R)$-gravity is that the 4-acceleration $A_a$ be given as a gradient of some scalar $\psi$ (see Eqn \eref{R5ir}). Upon specialising to dust-fluid cases, we have seen that no new constraints arise and hence the limiting field equations propagate consistently. But if one further specialises to shear-free dust, then we get a vanishing $H_{ab}$ and a temporally and spatially consistent constraint equation \eref{shearconst1}. We have shown also that the electric component of the Weyl tensor does not vanish as a result of the non-vanishing contribution of the anisotropic pressure from the curvature fluid. In addition, we have shown that, as a result of the shear-free condition, $q^R_a$ is irrotational and can be given as a gradient of some scalar function $\phi=\sfrac{2}{3}\Theta+C$ for some spatially constant $C$. 

We have also shown that the linearised field equations of $f(R)$-gravity in irrotational shear-free dust spacetimes with vanishing Weyl tensor as well as the purely gravito-magnetic spacetimes are consistent. In the case of purely radiative irrotational dust spacetimes, the consistency requirement implies that such models need not be homogeneous, unlike their GR counterparts.

Another subclass of irrotational spacetimes we looked at is the non-expanding case, where the new constraint \eref{ray2s} appears as a result of the $\Theta=0$ restriction. Dust spacetimes in this subclass are further constrained by Eqn \eref{fconstr} whereas general shear-free cases are constrained by Eqn \eref{sig2sc}. Moreover, the very special case of shear-free dust solutions results in $E_{ab}$ and $\pi^R_{ab}$ being constants over cosmic time as well as in the vanishing of $q^{R}_{a}$.

In short, we have explored some [sub]classes of irrotational cosmological models and shown how these models put restrictions on the possible forms of the underlying $f(R)$ gravitational theory.

\ack ME acknowledges the hospitality of the Astrophysics Group (Department of Physics, NWU-Mahikeng) during the preparation of most of this work. AA is supported by a NWU Postdoctoral fellowship.

\appendix

\section{Useful Linearised Differential Identities}
For all scalars $f$, vectors $V_a$ and tensors that vanish in the background,
$S_{ab}=S_{\la ab\ra}$, the following linearised identities hold \cite{ carloni08, maartens98, maartens97}:
\begin{eqnarray}
\left(\D_{\la a}\D_{b\ra}f\right)^{.}&=&\D_{\la a}\D_{b\ra}\dot{f}-\sfrac{2}{3}\Theta\D_{\la a}\D_{b\ra}f+\dot{f}\D_{\la a}A_{b\ra}\label{a0}\;,\\
\ep^{abc}\D_b \D_cf &=& 0 \label{a1}\;, \\
\ep_{cda}\D^{c}\D_{\la b}\D^{d\ra}f&=&\ep_{cda}\D^{c}\D_{( b}\D^{d)}f=\ep_{cda}\D^{c}\D_{ b}\D^{d}f=0\label{a2}\;,\\
\D^2\left(\D_af\right) &=&\D_a\left(\D^2f\right) 
+\sfrac{1}{3}\tl{R}\D_a f \label{a19}\;,\\
\left(\D_af\right)^{\rd} &=& \D_a\dot{f}-\sfrac{1}{3}\Theta\D_af+\dot{f}A_a 
\label{a14}\;,\\
\left(\D_aS_{b\cdots}\right)^{\rd} &=& \D_a\dot{S}_{b\cdots}
-\sfrac{1}{3}\Theta\D_aS_{b\cdots}
\label{a15}\;,\\
\left(\D^2 f\right)^{\rd} &=& \D^2\dot{f}-\sfrac{2}{3}\Theta\D^2 f 
+\dot{f}\D^a A_a \label{a21}\;,\\
\D_{[a}\D_{b]}V_c &=& 
-\sfrac{1}{6}\tl{R}V_{[a}h_{b]c} \label{a16}\;,\\
\D_{[a}\D_{b]}S^{cd} &=& -\sfrac{1}{3}\tl{R}S_{[a}{}^{(c}h_{b]}{}^{d)} \label{a17}\;,\\
\D^a\left(\ep_{abc}\D^bV^c\right) &=& 0 \label{a20}\;,\\
\label{divcurl}\D_b\left(\ep^{cd\la a}\D_c S^{b\ra}_d\right) &=& {\ts{1\over2}}\ep^{abc}\D_b \left(\D_d S^d_c\right)\;,\\
\text{curlcurl} V_{a}&=&\D_{a}\left(\D^{b}V_{b}\right)-\D^{2}V_{a}+\sfrac{2}{3}\left(\mu-\sfrac{1}{3}\Theta^{2}\right)V_{a}\label{curlcurla}\;,
\end{eqnarray}
where $\tl{R}\equiv 2\left(\mu-\frac13\Theta^2\right)$ is the 3-curvature scalar.
%
\section*{References}

\bibliography{bibliography}

\begin{thebibliography}{10}
\expandafter\ifx\csname url\endcsname\relax
  \def\url#1{\texttt{#1}}\fi
\expandafter\ifx\csname urlprefix\endcsname\relax\def\urlprefix{URL }\fi
\providecommand{\bibinfo}[2]{#2}
\providecommand{\eprint}[2][]{\url{#2}}

\bibitem{sotiriou10}
\bibinfo{author}{Sotiriou, T.~P.} \& \bibinfo{author}{Faraoni, V.}
\newblock \bibinfo{title}{${f(R)}$ theories of gravity}.
\newblock \emph{\bibinfo{journal}{Reviews of Modern Physics}}
  \textbf{\bibinfo{volume}{82}}, \bibinfo{pages}{451} (\bibinfo{year}{2010}).

\bibitem{clifton12}
\bibinfo{author}{Clifton, T.}, \bibinfo{author}{Ferreira, P.~G.},
  \bibinfo{author}{Padilla, A.} \& \bibinfo{author}{Skordis, C.}
\newblock \bibinfo{title}{Modified gravity and cosmology}.
\newblock \emph{\bibinfo{journal}{Physics Reports}}
  \textbf{\bibinfo{volume}{513}}, \bibinfo{pages}{1--189}
  (\bibinfo{year}{2012}).

\bibitem{capozziello11extended}
\bibinfo{author}{Capozziello, S.} \& \bibinfo{author}{De~Laurentis, M.}
\newblock \bibinfo{title}{Extended theories of gravity}.
\newblock \emph{\bibinfo{journal}{Physics Reports}}
  \textbf{\bibinfo{volume}{509}}, \bibinfo{pages}{167--321}
  (\bibinfo{year}{2011}).

\bibitem{modesto12}
\bibinfo{author}{Modesto, L.}
\newblock \bibinfo{title}{Super-renormalizable quantum gravity}.
\newblock \emph{\bibinfo{journal}{Phys. Rev. D}} \textbf{\bibinfo{volume}{86}},
  \bibinfo{pages}{044005} (\bibinfo{year}{2012}).

\bibitem{biswas12}
\bibinfo{author}{Biswas, T.}, \bibinfo{author}{Gerwick, E.},
  \bibinfo{author}{Koivisto, T.} \& \bibinfo{author}{Mazumdar, A.}
\newblock \bibinfo{title}{{Towards singularity and ghost free theories of
  gravity}}.
\newblock \emph{\bibinfo{journal}{Phys.Rev.Lett.}}
  \textbf{\bibinfo{volume}{108}}, \bibinfo{pages}{031101}
  (\bibinfo{year}{2012}).

\bibitem{nojiri03}
\bibinfo{author}{Nojiri, S.} \& \bibinfo{author}{Odintsov, S.~D.}
\newblock \bibinfo{title}{Modified gravity with negative and positive powers of
  curvature: Unification of inflation and cosmic acceleration}.
\newblock \emph{\bibinfo{journal}{Physical Review D}}
  \textbf{\bibinfo{volume}{68}}, \bibinfo{pages}{123512}
  (\bibinfo{year}{2003}).

\bibitem{staro80}
\bibinfo{author}{{Starobinsky}, A.~A.}
\newblock \bibinfo{title}{{A new type of isotropic cosmological models without
  singularity}}.
\newblock \emph{\bibinfo{journal}{Physics Letters B}}
  \textbf{\bibinfo{volume}{91}}, \bibinfo{pages}{99--102}
  (\bibinfo{year}{1980}).

\bibitem{carroll04}
\bibinfo{author}{Carroll, S.}, \bibinfo{author}{Duvvuri, V.},
  \bibinfo{author}{Turner, M.} \& \bibinfo{author}{Trodden, M.}
\newblock \bibinfo{title}{Is cosmic speed-up due to new gravitational physics?}
\newblock \emph{\bibinfo{journal}{Physical Review D}}
  \textbf{\bibinfo{volume}{70}}, \bibinfo{pages}{043528}.

\bibitem{faraoni08}
\bibinfo{author}{Faraoni, V.}
\newblock \bibinfo{title}{${f(R)}$ gravity: successes and challenges}.
\newblock \emph{\bibinfo{journal}{arXiv preprint arXiv:0810.2602}}
  (\bibinfo{year}{2008}).

\bibitem{sotiriou07}
\bibinfo{author}{Sotiriou, T.~P.} \& \bibinfo{author}{Liberati, S.}
\newblock \bibinfo{title}{Metric-affine ${f(R)}$ theories of gravity}.
\newblock \emph{\bibinfo{journal}{Annals of Physics}}
  \textbf{\bibinfo{volume}{322}}, \bibinfo{pages}{935--966}
  (\bibinfo{year}{2007}).

\bibitem{nojiri2011unified}
\bibinfo{author}{Nojiri, S.} \& \bibinfo{author}{Odintsov, S.~D.}
\newblock \bibinfo{title}{Unified cosmic history in modified gravity: from
  ${f(R)}$ theory to lorentz non-invariant models}.
\newblock \emph{\bibinfo{journal}{Physics Reports}}
  \textbf{\bibinfo{volume}{505}}, \bibinfo{pages}{59--144}
  (\bibinfo{year}{2011}).

\bibitem{nojiri06}
\bibinfo{author}{Nojiri, S.} \& \bibinfo{author}{Odintsov, S.~D.}
\newblock \bibinfo{title}{Modified ${f(R)}$ gravity consistent with realistic
  cosmology: From a matter dominated epoch to a dark energy universe}.
\newblock \emph{\bibinfo{journal}{Physical Review D}}
  \textbf{\bibinfo{volume}{74}}, \bibinfo{pages}{086005}
  (\bibinfo{year}{2006}).

\bibitem{nojiri07}
\bibinfo{author}{Nojiri, S.} \& \bibinfo{author}{Odintsov, S.~D.}
\newblock \bibinfo{title}{Unifying inflation with ${\Lambda}${CDM} epoch in
  modified ${f(R)}$ gravity consistent with solar system tests}.
\newblock \emph{\bibinfo{journal}{Physics Letters B}}
  \textbf{\bibinfo{volume}{657}}, \bibinfo{pages}{238--245}
  (\bibinfo{year}{2007}).

\bibitem{sriva08}
\bibinfo{author}{Srivastava, S.}
\newblock \bibinfo{title}{Curvature inspired cosmological scenario}.
\newblock \emph{\bibinfo{journal}{International Journal of Theoretical
  Physics}} \textbf{\bibinfo{volume}{47}}, \bibinfo{pages}{1966--1978}
  (\bibinfo{year}{2008}).

\bibitem{capoz06}
\bibinfo{author}{Capozziello, S.}, \bibinfo{author}{Cardone, V.} \&
  \bibinfo{author}{Troisi, A.}
\newblock \bibinfo{title}{Dark energy and dark matter as curvature effects?}
\newblock \emph{\bibinfo{journal}{Journal of Cosmology and Astroparticle
  Physics}} \textbf{\bibinfo{volume}{2006}}, \bibinfo{pages}{001}
  (\bibinfo{year}{2006}).

\bibitem{de2010f}
\bibinfo{author}{De~Felice, A.} \& \bibinfo{author}{Tsujikawa, S.}
\newblock \bibinfo{title}{${f(R)}$ theories}.
\newblock \emph{\bibinfo{journal}{Living Rev. Rel}}
  \textbf{\bibinfo{volume}{13}}, \bibinfo{pages}{1002--4928}
  (\bibinfo{year}{2010}).

\bibitem{magnano87}
\bibinfo{author}{Magnano, G.}, \bibinfo{author}{Ferraris, M.} \&
  \bibinfo{author}{Francaviglia, M.}
\newblock \bibinfo{title}{Nonlinear gravitational lagrangians}.
\newblock \emph{\bibinfo{journal}{General relativity and gravitation}}
  \textbf{\bibinfo{volume}{19}}, \bibinfo{pages}{465--479}
  (\bibinfo{year}{1987}).

\bibitem{rippl96}
\bibinfo{author}{Rippl, S.}, \bibinfo{author}{van Elst, H.},
  \bibinfo{author}{Tavakol, R.} \& \bibinfo{author}{Taylor, D.}
\newblock \bibinfo{title}{Kinematics and dynamics of ${f (R)}$ theories of
  gravity}.
\newblock \emph{\bibinfo{journal}{General Relativity and Gravitation}}
  \textbf{\bibinfo{volume}{28}}, \bibinfo{pages}{193--205}
  (\bibinfo{year}{1996}).

\bibitem{amendola07}
\bibinfo{author}{Amendola, L.}, \bibinfo{author}{Polarski, D.} \&
  \bibinfo{author}{Tsujikawa, S.}
\newblock \bibinfo{title}{Are ${f(R)}$ dark energy models cosmologically
  viable?}
\newblock \emph{\bibinfo{journal}{Physical review letters}}
  \textbf{\bibinfo{volume}{98}}, \bibinfo{pages}{131302}
  (\bibinfo{year}{2007}).

\bibitem{CDCT05}
\bibinfo{author}{Carloni, S.}, \bibinfo{author}{Dunsby, P.~K.},
  \bibinfo{author}{Capozziello, S.} \& \bibinfo{author}{Troisi, A.}
\newblock \bibinfo{title}{Cosmological dynamics of ${R^n}$ gravity}.
\newblock \emph{\bibinfo{journal}{Classical and Quantum Gravity}}
  \textbf{\bibinfo{volume}{22}}, \bibinfo{pages}{4839} (\bibinfo{year}{2005}).

\bibitem{carloni2009}
\bibinfo{author}{Carloni, S.}, \bibinfo{author}{Troisi, A.} \&
  \bibinfo{author}{Dunsby, P.}
\newblock \bibinfo{title}{Some remarks on the dynamical systems approach to
  fourth order gravity}.
\newblock \emph{\bibinfo{journal}{General Relativity and Gravitation}}
  \textbf{\bibinfo{volume}{41}}, \bibinfo{pages}{1757--1776}
  (\bibinfo{year}{2009}).

\bibitem{leach06}
\bibinfo{author}{Leach, J.~A.}, \bibinfo{author}{Carloni, S.} \&
  \bibinfo{author}{Dunsby, P.~K.}
\newblock \bibinfo{title}{Shear dynamics in {Bianchi I} cosmologies with
  ${R^n}$-gravity}.
\newblock \emph{\bibinfo{journal}{Classical and Quantum Gravity}}
  \textbf{\bibinfo{volume}{23}}, \bibinfo{pages}{4915} (\bibinfo{year}{2006}).

\bibitem{goheer2009power}
\bibinfo{author}{Goheer, N.}, \bibinfo{author}{Larena, J.} \&
  \bibinfo{author}{Dunsby, P.~K.}
\newblock \bibinfo{title}{Power-law cosmic expansion in ${f(R)}$ gravity
  models}.
\newblock \emph{\bibinfo{journal}{Physical Review D}}
  \textbf{\bibinfo{volume}{80}}, \bibinfo{pages}{61301} (\bibinfo{year}{2009}).

\bibitem{dunsby10}
\bibinfo{author}{Dunsby, P.~K.}, \bibinfo{author}{Elizalde, E.},
  \bibinfo{author}{Goswami, R.}, \bibinfo{author}{Odintsov, S.} \&
  \bibinfo{author}{Saez-Gomez, D.}
\newblock \bibinfo{title}{${\Lambda}${CDM} universe in ${f(R)}$ gravity}.
\newblock \emph{\bibinfo{journal}{Physical Review D}}
  \textbf{\bibinfo{volume}{82}}, \bibinfo{pages}{023519}
  (\bibinfo{year}{2010}).

\bibitem{bean07}
\bibinfo{author}{Bean, R.}, \bibinfo{author}{Bernat, D.},
  \bibinfo{author}{Pogosian, L.}, \bibinfo{author}{Silvestri, A.} \&
  \bibinfo{author}{Trodden, M.}
\newblock \bibinfo{title}{Dynamics of linear perturbations in ${f(R)}$
  gravity}.
\newblock \emph{\bibinfo{journal}{Physical Review D}}
  \textbf{\bibinfo{volume}{75}}, \bibinfo{pages}{064020}
  (\bibinfo{year}{2007}).

\bibitem{song07}
\bibinfo{author}{Song, Y.-S.}, \bibinfo{author}{Hu, W.} \&
  \bibinfo{author}{Sawicki, I.}
\newblock \bibinfo{title}{Large scale structure of ${f(R)}$ gravity}.
\newblock \emph{\bibinfo{journal}{Physical Review D}}
  \textbf{\bibinfo{volume}{75}}, \bibinfo{pages}{044004}
  (\bibinfo{year}{2007}).

\bibitem{abebe12}
\bibinfo{author}{Abebe, A.}, \bibinfo{author}{Abdelwahab, M.},
  \bibinfo{author}{de~la Cruz-Dombriz, {\'A}.} \& \bibinfo{author}{Dunsby,
  P.~K.}
\newblock \bibinfo{title}{Covariant gauge-invariant perturbations in multifluid
  ${f(R)}$ gravity}.
\newblock \emph{\bibinfo{journal}{Classical and Quantum Gravity}}
  \textbf{\bibinfo{volume}{29}}, \bibinfo{pages}{135011}
  (\bibinfo{year}{2012}).

\bibitem{carloni08}
\bibinfo{author}{Carloni, S.}, \bibinfo{author}{Dunsby, P.} \&
  \bibinfo{author}{Troisi, A.}
\newblock \bibinfo{title}{Evolution of density perturbations in {${f(R)}$}
  gravity}.
\newblock \emph{\bibinfo{journal}{Physical Review D}}
  \textbf{\bibinfo{volume}{77}}, \bibinfo{pages}{024024}
  (\bibinfo{year}{2008}).

\bibitem{ananda09}
\bibinfo{author}{Ananda, K.~N.}, \bibinfo{author}{Carloni, S.} \&
  \bibinfo{author}{Dunsby, P.~K.}
\newblock \bibinfo{title}{Structure growth in ${f(R)}$ theories of gravity with
  a dust equation of state}.
\newblock \emph{\bibinfo{journal}{Classical and Quantum Gravity}}
  \textbf{\bibinfo{volume}{26}}, \bibinfo{pages}{235018}
  (\bibinfo{year}{2009}).

\bibitem{abebe13}
\bibinfo{author}{Abebe, A.}, \bibinfo{author}{de~la Cruz-Dombriz, {\'A}.} \&
  \bibinfo{author}{Dunsby, P.~K.}
\newblock \bibinfo{title}{Large scale structure constraints for a class of ${f
  (R)}$ theories of gravity}.
\newblock \emph{\bibinfo{journal}{Physical Review D}}
  \textbf{\bibinfo{volume}{88}}, \bibinfo{pages}{044050}
  (\bibinfo{year}{2013}).

\bibitem{ananda08}
\bibinfo{author}{Ananda, K.~N.}, \bibinfo{author}{Carloni, S.} \&
  \bibinfo{author}{Dunsby, P.~K.}
\newblock \bibinfo{title}{A detailed analysis of structure growth in ${f(R)}$
  theories of gravity}.
\newblock \emph{\bibinfo{journal}{arXiv preprint arXiv:0809.3673}}
  (\bibinfo{year}{2008}).

\bibitem{abebe14b}
\bibinfo{author}{Abebe, A.}
\newblock \bibinfo{title}{{Breaking the cosmological background degeneracy by
  two-fluid perturbations in ${f(R)}$ gravity}}.
\newblock \emph{\bibinfo{journal}{Int.J.Mod.Phys. D}}
  \textbf{\bibinfo{volume}{24}}, \bibinfo{pages}{1550053}.

\bibitem{dombriz08}
\bibinfo{author}{de~La~Cruz-Dombriz, A.}, \bibinfo{author}{Dobado, A.} \&
  \bibinfo{author}{Maroto, A.}
\newblock \bibinfo{title}{Evolution of density perturbations in ${f(R)}$
  theories of gravity}.
\newblock \emph{\bibinfo{journal}{Physical Review D}}
  \textbf{\bibinfo{volume}{77}}, \bibinfo{pages}{123515}
  (\bibinfo{year}{2008}).

\bibitem{matarrese94}
\bibinfo{author}{Matarrese, S.}, \bibinfo{author}{Pantano, O.} \&
  \bibinfo{author}{Saez, D.}
\newblock \bibinfo{title}{General relativistic dynamics of irrotational dust:
  Cosmological implications}.
\newblock \emph{\bibinfo{journal}{Physical review letters}}
  \textbf{\bibinfo{volume}{72}}, \bibinfo{pages}{320--323}
  (\bibinfo{year}{1994}).

\bibitem{bert94}
\bibinfo{author}{Bertschinger, E.} \& \bibinfo{author}{Jain, B.}
\newblock \bibinfo{title}{Gravitational instability of cold matter}.
\newblock \emph{\bibinfo{journal}{The Astrophysical Journal}}
  \textbf{\bibinfo{volume}{431}}, \bibinfo{pages}{486--494}
  (\bibinfo{year}{1994}).

\bibitem{maartens98}
\bibinfo{author}{Maartens, R.}
\newblock \bibinfo{title}{Covariant velocity and density perturbations in
  quasi-newtonian cosmologies}.
\newblock \emph{\bibinfo{journal}{Physical Review D}}
  \textbf{\bibinfo{volume}{58}}, \bibinfo{pages}{124006}
  (\bibinfo{year}{1998}).

\bibitem{wyll06}
\bibinfo{author}{Wylleman, L.}
\newblock \bibinfo{title}{Anti-newtonian universes do not exist}.
\newblock \emph{\bibinfo{journal}{Classical and Quantum Gravity}}
  \textbf{\bibinfo{volume}{23}}, \bibinfo{pages}{2727} (\bibinfo{year}{2006}).

\bibitem{maartens94}
\bibinfo{author}{Maartens, R.} \& \bibinfo{author}{Matravers, D.}
\newblock \bibinfo{title}{Isotropic and semi-isotropic observations in
  cosmology}.
\newblock \emph{\bibinfo{journal}{Classical and Quantum Gravity}}
  \textbf{\bibinfo{volume}{11}}, \bibinfo{pages}{2693} (\bibinfo{year}{1994}).

\bibitem{ellis85id}
\bibinfo{author}{Ellis, G.}, \bibinfo{author}{Nel, S.},
  \bibinfo{author}{Maartens, R.}, \bibinfo{author}{Stoeger, W.} \&
  \bibinfo{author}{Whitman, A.}
\newblock \bibinfo{title}{Ideal observational cosmology}.
\newblock \emph{\bibinfo{journal}{Physics Reports}}
  \textbf{\bibinfo{volume}{124}}, \bibinfo{pages}{315--417}
  (\bibinfo{year}{1985}).

\bibitem{ellis67}
\bibinfo{author}{Ellis, G.}
\newblock \bibinfo{title}{Dynamics of pressure-free matter in general
  relativity}.
\newblock \emph{\bibinfo{journal}{Journal of Mathematical Physics}}
  \textbf{\bibinfo{volume}{8}}, \bibinfo{pages}{1171} (\bibinfo{year}{1967}).

\bibitem{Ellis98}
\bibinfo{author}{Ellis, G.} \& \bibinfo{author}{van Elst, H.}
\newblock \bibinfo{title}{Cosmological models}.
\newblock In \emph{\bibinfo{booktitle}{Theoretical and Observational
  Cosmology}}, \bibinfo{pages}{1--116} (\bibinfo{publisher}{Dordrecht: Kluver},
  \bibinfo{year}{1999}).

\bibitem{maartens97}
\bibinfo{author}{Maartens, R.} \& \bibinfo{author}{Triginer, J.}
\newblock \bibinfo{title}{Density perturbations with relativistic
  thermodynamics}.
\newblock \emph{\bibinfo{journal}{Physical Review D}}
  \textbf{\bibinfo{volume}{56}}, \bibinfo{pages}{4640} (\bibinfo{year}{1997}).

\bibitem{van97}
\bibinfo{author}{van Elst, H.}, \bibinfo{author}{Uggla, C.},
  \bibinfo{author}{Lesame, W.~M.}, \bibinfo{author}{Ellis, G.} \&
  \bibinfo{author}{Maartens, R.}
\newblock \bibinfo{title}{Integrability of irrotational silent cosmological
  models}.
\newblock \emph{\bibinfo{journal}{Classical and Quantum Gravity}}
  \textbf{\bibinfo{volume}{14}}, \bibinfo{pages}{1151} (\bibinfo{year}{1997}).

\bibitem{lesame95}
\bibinfo{author}{Lesame, W.}, \bibinfo{author}{Dunsby, P.} \&
  \bibinfo{author}{Ellis, G.}
\newblock \bibinfo{title}{Integrability conditions for irrotational dust with a
  purely electric weyl tensor: A tetrad analysis}.
\newblock \emph{\bibinfo{journal}{Physical Review D}}
  \textbf{\bibinfo{volume}{52}}, \bibinfo{pages}{3406} (\bibinfo{year}{1995}).

\bibitem{elst98}
\bibinfo{author}{van Elst, H.} \& \bibinfo{author}{Ellis, G.}
\newblock \bibinfo{title}{Quasi-newtonian dust cosmologies}.
\newblock \emph{\bibinfo{journal}{Classical and Quantum Gravity}}
  \textbf{\bibinfo{volume}{15}}, \bibinfo{pages}{3545} (\bibinfo{year}{1998}).

\bibitem{macc98}
\bibinfo{author}{MacCallum, M.}
\newblock \bibinfo{title}{Integrability in tetrad formalisms and conservation
  in cosmology}.
\newblock \emph{\bibinfo{journal}{arXiv preprint gr-qc/9806003}}
  (\bibinfo{year}{1998}).

\bibitem{buchdal70}
\bibinfo{author}{{Buchdahl}, H.~A.}
\newblock \bibinfo{title}{{Non-linear Lagrangians and cosmological theory}}.
\newblock \emph{\bibinfo{journal}{Monthly Notices of the Royal Astronomical
  Society}} \textbf{\bibinfo{volume}{150}}, \bibinfo{pages}{1}
  (\bibinfo{year}{1970}).

\bibitem{Abebe2011}
\bibinfo{author}{Abebe, A.}, \bibinfo{author}{Goswami, R.} \&
  \bibinfo{author}{Dunsby, P.}
\newblock \bibinfo{title}{{Shear-free perturbations of ${f(R)}$ gravity}}.
\newblock \emph{\bibinfo{journal}{Physical Review D}} \bibinfo{pages}{1--7}.

\bibitem{betschart}
\bibinfo{author}{Betschart, G.}
\newblock \emph{\bibinfo{title}{{General relativistic electrodynamics with
  applicantions in cosmology and astrophysics}}}.
\newblock Ph.D. thesis, \bibinfo{school}{University of Cape Town}
  (\bibinfo{year}{2005}).

\bibitem{SW74}
\bibinfo{author}{Stewart, J.} \& \bibinfo{author}{Walker, M.}
\newblock \bibinfo{title}{Perturbations of space-times in general relativity}.
\newblock \emph{\bibinfo{journal}{Proceedings of the Royal Society of London.
  A. Mathematical and Physical Sciences}} \textbf{\bibinfo{volume}{341}},
  \bibinfo{pages}{49--74} (\bibinfo{year}{1974}).

\bibitem{ellis89}
\bibinfo{author}{Ellis, G.} \& \bibinfo{author}{Bruni, M.}
\newblock \bibinfo{title}{Covariant and gauge-invariant approach to
  cosmological density fluctuations}.
\newblock \emph{\bibinfo{journal}{Physical Review D}}
  \textbf{\bibinfo{volume}{40}}, \bibinfo{pages}{1804} (\bibinfo{year}{1989}).

\bibitem{ellis12}
\bibinfo{author}{Ellis, G.}, \bibinfo{author}{Maartens, R.} \&
  \bibinfo{author}{MacCallum, M.~A.}
\newblock \emph{\bibinfo{title}{Relativistic cosmology}}
  (\bibinfo{publisher}{Cambridge University Press}, \bibinfo{year}{2012}).

\bibitem{godel52}
\bibinfo{author}{G{\"o}del, K.}
\newblock \bibinfo{title}{Rotating universes in general relativity theory}.
\newblock In \emph{\bibinfo{booktitle}{Proceedings of the International
  Congress of Mathematicians Edited by LM Graves et al., Cambridge, Mass. 1952,
  vol. 1, p. 175.}}, vol.~\bibinfo{volume}{1}, \bibinfo{pages}{175}
  (\bibinfo{year}{1952}).

\bibitem{goldberg62}
\bibinfo{author}{Goldberg, J.} \& \bibinfo{author}{Sachs, R.}
\newblock \bibinfo{title}{A theorem on {Petrov} type (field equations for
  proving theorem identifying geometrical properties of null congruence with
  existence of algebraically special {Riemann} tensor)}.
\newblock \emph{\bibinfo{journal}{Acta Physica Polonica}}
  \textbf{\bibinfo{volume}{22}}, \bibinfo{pages}{13--23}
  (\bibinfo{year}{1962}).

\bibitem{robinson63}
\bibinfo{author}{Robinson, I.} \& \bibinfo{author}{Schild, A.}
\newblock \bibinfo{title}{Generalization of a theorem by goldberg and sachs}.
\newblock \emph{\bibinfo{journal}{Journal of Mathematical Physics}}
  \textbf{\bibinfo{volume}{4}}, \bibinfo{pages}{484} (\bibinfo{year}{1963}).

\bibitem{abebe2013}
\bibinfo{author}{Abebe, A.}
\newblock \emph{\bibinfo{title}{Beyond concordance cosmology}}.
\newblock Ph.D. thesis, \bibinfo{school}{UCT (University of Cape Town)}
  (\bibinfo{year}{2013}).

\bibitem{ellis2011S}
\bibinfo{author}{Ellis, G.}
\newblock \bibinfo{title}{Shear free solutions in general relativity theory}.
\newblock \emph{\bibinfo{journal}{General Relativity and Gravitation}}
  \textbf{\bibinfo{volume}{43}}, \bibinfo{pages}{3253--3268}
  (\bibinfo{year}{2011}).

\bibitem{narlikar99}
\bibinfo{author}{Narlikar, J.~V.}
\newblock \bibinfo{title}{Spinning universes in newtonian cosmology}.
\newblock In \emph{\bibinfo{booktitle}{On Einstein's Path}},
  \bibinfo{pages}{319--327} (\bibinfo{publisher}{Springer},
  \bibinfo{year}{1999}).

\bibitem{narlikar63}
\bibinfo{author}{Narlikar, J.}
\newblock \bibinfo{title}{Newtonian universes with shear and rotation}.
\newblock \emph{\bibinfo{journal}{Monthly Notices of the Royal Astronomical
  Society}} \textbf{\bibinfo{volume}{126}}, \bibinfo{pages}{203}
  (\bibinfo{year}{1963}).

\bibitem{lesame96}
\bibinfo{author}{Lesame, W.}, \bibinfo{author}{Ellis, G.} \&
  \bibinfo{author}{Dunsby, P.}
\newblock \bibinfo{title}{Irrotational dust with divh= 0}.
\newblock \emph{\bibinfo{journal}{Physical Review D}}
  \textbf{\bibinfo{volume}{53}}, \bibinfo{pages}{738} (\bibinfo{year}{1996}).

\bibitem{maar97c}
\bibinfo{author}{Maartens, R.}, \bibinfo{author}{Lesame, W.~M.} \&
  \bibinfo{author}{Ellis, G.}
\newblock \bibinfo{title}{Consistency of dust solutions with div h= 0}.
\newblock \emph{\bibinfo{journal}{Physical Review D}}
  \textbf{\bibinfo{volume}{55}}, \bibinfo{pages}{5219} (\bibinfo{year}{1997}).

\bibitem{maartens97lin}
\bibinfo{author}{Maartens, R.}
\newblock \bibinfo{title}{Linearization instability of gravity waves?}
\newblock \emph{\bibinfo{journal}{Physical Review D}}
  \textbf{\bibinfo{volume}{55}}, \bibinfo{pages}{463} (\bibinfo{year}{1997}).

\bibitem{van07}
\bibinfo{author}{Van~den Bergh, N.}, \bibinfo{author}{Bastiaensen, B.},
  \bibinfo{author}{Karimian, H.} \& \bibinfo{author}{Wylleman, L.}
\newblock \bibinfo{title}{Purely radiative irrotational dust spacetimes}.
\newblock In \emph{\bibinfo{booktitle}{Journal of Physics: Conference Series}},
  vol.~\bibinfo{volume}{66}, \bibinfo{pages}{012023}
  (\bibinfo{organization}{IOP Publishing}, \bibinfo{year}{2007}).

\bibitem{maartens1998}
\bibinfo{author}{Maartens, R.}, \bibinfo{author}{Lesame, W.~M.} \&
  \bibinfo{author}{Ellis, G.}
\newblock \bibinfo{title}{Newtonian-like and anti-newtonian universes}.
\newblock \emph{\bibinfo{journal}{Classical and Quantum Gravity}}
  \textbf{\bibinfo{volume}{15}}, \bibinfo{pages}{1005} (\bibinfo{year}{1998}).

\bibitem{tomita72}
\bibinfo{author}{Tomita, K.}
\newblock \bibinfo{title}{Primordial irregularities in the early universe}.
\newblock \emph{\bibinfo{journal}{Progress of Theoretical Physics}}
  \textbf{\bibinfo{volume}{48}}, \bibinfo{pages}{1503--1516}
  (\bibinfo{year}{1972}).

\bibitem{tomita75}
\bibinfo{author}{Tomita, K.}
\newblock \bibinfo{title}{Evolution of irregularities in a chaotic early
  universe}.
\newblock \emph{\bibinfo{journal}{Progress of Theoretical Physics}}
  \textbf{\bibinfo{volume}{54}}, \bibinfo{pages}{730--739}
  (\bibinfo{year}{1975}).

\bibitem{amare14a}
\bibinfo{author}{Abebe, A.}
\newblock \bibinfo{title}{Anti-newtonian cosmologies in ${f(R)}$ gravity}.
\newblock \emph{\bibinfo{journal}{Classical and Quantum Gravity}}
  \textbf{\bibinfo{volume}{31}}, \bibinfo{pages}{115011}
  (\bibinfo{year}{2014}).

\end{thebibliography}
\bibliographystyle{naturemag}

\end{document}